# Potential of Ozone Formation by the Smog Mechanism to shield the surface of the Early Earth from UV radiation?

John Lee Grenfell,[1*] Barbara Stracke,[1] Beate Patzer,[2] Ruth Titz[1] and Heike Rauer[1]

(1) Institute of Planetary Research
German Aerospace Centre (DLR)
Rutherford Str. 2
12489 Berlin

(2) Centre for Astronomy and Astrophysics
Technical University Berlin (TUB)
Hardenbergstr. 36
10623 Berlin

*Please address correspondence to: John Lee Grenfell
Institute of Planetary Research
German Aerospace Centre (DLR)
Rutherford Str. 2
12489 Berlin
phone: +49 30 67055 337
fax: +40 30 67055 340
lee.grenfell@dlr.de

*Abstract: We propose that the photochemical smog mechanism produced substantial ozone ($O_3$) in the troposphere during the Proterozoic, which contributed to ultraviolet (UV) radiation shielding hence favoured the establishment of life. The smog mechanism is well-established and is associated with pollution hazes which sometimes cover modern cities. The mechanism proceeds via the oxidation of volatile organic compounds (VOCs) such as methane ($CH_4$) in the presence of UV radiation and nitrogen oxides ($NO_x$). It would have been particularly favoured during the Proterozoic given the high levels of $CH_4$ (up to 1000 ppm) recently suggested. Proterozoic UV levels on the surface of the Earth were generally higher compared with today, which would also have favoured the mechanism. On the other hand, Proterozoic $O_2$ required in the final step of the smog mechanism to form $O_3$ was less abundant compared with present times. Further, results are sensitive to Proterozoic $NO_x$ concentrations, which are challenging to predict, since they depend on uncertain quantities such as $NO_x$ source emissions and OH concentrations. We review $NO_x$ sources during the Proterozoic and apply a photochemical box model having methane oxidation with $NO_x$, $HO_x$ and $O_x$ chemistry to estimate the $O_3$ production from the smog mechanism. Runs suggest the smog mechanism during the Proterozoic can produce about double present day ozone columns for $NO_x$ levels of 1.53 $10^{-9}$ by volume mixing ratio, which was attainable according to our $NO_x$ source analysis, with 1% present atmospheric levels (PALs) of $O_2$. Clearly, forming ozone in the troposphere is a trade-off for survivability – on the one hand harmful UV is blocked, but on the other hand ozone is a respiratory irratant, which becomes fatal at concentrations exceeding about 1 ppmv.*

*Key words: early Earth, UV shielding, ozone, smog mechanism*

## 1. Introduction

Understanding the atmospheric composition of the early Earth is linked to fundamental questions concerning the origin and establishment of life (e.g. *Bada*, 2004). The chemistry of the early Earth atmosphere is also highly relevant to forthcoming missions such as Darwin (Leger *et al.*, 1996; Leger, 2000) and NASA's Terrestrial Planet Finder (TPF) (Beichman *et al.*, 1999) which will search for biomarker molecules on 'earthlike' worlds, because many stars in the solar neighbourhood (<30 pc) relevant to Darwin are younger than our sun (Lammer *et al.*, 2005).

Several UV-shielding mechanisms have been proposed to explain how living organisms managed to colonise the surface despite strong UV exposure in the early stages of earth history. For example, Sagan and Chyba (1997) proposed UV-shielding from a fine, hydrocarbon haze, beneath which high levels of ammonia could accumulate. Pavlov *et al.* (2001) however, could not reproduce their result, instead calculating a much weaker shielding effect, attributed to differences in the aerosol size distribution. Selsis *et al.* (2002a) suggested some abiotic stratospheric ozone production hence UV-shielding via carbon dioxide ($CO_2$) and water ($H_2O$) photolysis. Cleaves and Miller (1998) proposed that shielding was provided by an “oil slick” of organic material which covered the early oceans. Thomas *et al.* (2004) investigated abiotic $O_3$ production hence UV shielding in a $CO_2$-rich atmosphere.

Similar to our work, Segura *et al.* (2003) considered $O_2$ and $O_3$ changes using a model including $CH_4$ oxidation extending to the troposphere. However, unlike this work, they did not calculate large tropospheric $O_3$ changes, presumably because they adopted lower $NO_x$ abundances in their model[1]. Levine *et al.* (1979) studied the early Earth with a troposphere-stratosphere column model. They assumed 3 ppbv $NO_x$ i.e. comparable with our study. Kasting and Donahue (1980) suggested that 3 ppbv may be too high. Since then, additional $NO_x$ sources e.g. via cosmic rays have been better quantified, as we will discuss. Also, the Levine study did not include $CH_4$ oxidation and was published before the recently proposed high $CH_4$ values, so the smog mechanism operated much more slowly in their model. In this study, we suggest a link between tropospheric $O_3$ formed from the smog mechanism, UV shielding and the development of life on the early Earth.

[1]They assumed 21pptv $NO_x$ at the surface and 159 pptv $NO_x$ in the mid troposphere.

We have performed sensitivity studies with a photochemical model investigating the atmospheric composition of the early Earth. We varied the initial concentrations of key species which can influence the smog mechanism ($CH_4$, $O_2$, $NO_x$, $H_2$, CO) within currently reported uncertainty limits. An overview of Proterozoic $NO_x$ sources is provided, hence an ambient $NO_x$ concentration estimated for the early Earth. Finally, we estimate an overhead $O_3$ column produced via the smog mechanism, hence discuss its potential to provide an UV shield on the early Earth and the implications for life.

## 2. The Model

### 2.1 The smog mechanism and its chemical key species

The $O_3$ smog mechanism was first discussed in detail by Haagen-Smit *et al.* (1952) who studied smog formation in Los Angeles. The mechanism requires a volatile organic compound (VOC) such as $CH_4$ and some $NO_x$ in the presence of sunlight in order to form ozone, as illustrated below:

| | |
|---|---|
| $CH_4+OH \rightarrow CH_3+H_2O$ | (usually) rate determining i.e. slowest step |
| $CH_3+O_2+M \rightarrow CH_3O_2+M$ | fast formation of peroxy radical |
| $NO+CH_3O_2 \rightarrow NO_2+CH_3O$ | peroxy radical gives up an O to NO |
| $CH_3O+O_2 \rightarrow CH_2O+HO_2$ | $CH_3O$ rearranges to release $HO_2$ |
| $NO+HO_2 \rightarrow NO_2+OH$ | fast re-partitioning of $HO_x$ and $NO_x$ |
| $NO_2+h\nu \rightarrow NO+O$ | fast photolysis of $NO_2$ |
| $O+O_2+M \rightarrow O_3+M$ | three-body formation of ozone |

The mechanism begins with attack of hydroxyl (OH) upon a volatile organic compound such as methane, to form an organic radical (R) e.g. $CH_3$ as shown above plus water. Then, $O_2$ molecules quickly add on to R, to produce a peroxy radical ($RO_2$). This rapidly donates one of its O atoms to nitrogen monoxide (NO), to form nitrogen dioxide ($NO_2$) plus $CH_3O$. The $CH_3O$ undergoes a re-arrangement reaction and expels $HO_2$. The $HO_2$ quickly donates an O to NO, forming $NO_2$. The $NO_2$ quickly photolyses to reform the original NO plus O. The latter reacts with $O_2$ to form $O_3$. The smog mechanism has been confirmed by numerous modelling and theoretical studies e.g. (Demerjian *et al.*, 1974; Kumar and Mohan, 2002). Enhanced UV in the lower atmosphere during the Proterozoic clearly favours the smog mechanism. How much more surface UV would we expect without an $O_3$ shield? Modern day surface UV-B on Antarctica increases by a factor of three to six in September/October (Stamnes *et al.*, 1992) when up to half the overhead $O_3$ is destroyed due to the "$O_3$ hole". Segura *et al.* (2003) reported modelled surface UV-B increases by a factor of eleven for abiotic ($O_2$=$10^{-5}$ PAL) conditions. Clearly the abiotic atmosphere may have seen very harsh surface UV levels, despite the fainter sun, unless of course there featured a non-$O_3$ UV shield.

High $CH_4$ concentrations during the Proterozoic would clearly favour the smog mechanism. Evidence for high Proterozoic $CH_4$ is linked with the "faint young sun paradox". The early sun showed a weaker flux in the UV/visible range. This evokes high levels of greenhouse gases in the atmosphere to explain why liquid $H_2O$ existed on the early Earth, as suggested by the geologic record (e.g. Sagan and Mullen, 1972). Kasting (1993), for example, suggested high Proterozoic levels of $CO_2$ to explain the paradox. Paleosol data (Rye *et al.*, 1995) however, did not subsequently support such levels. Pavlov *et al.* (2000) then suggested that the paradox provides indirect evidence for high Proterozoic $CH_4$. Pavlov *et al.* (2000) and Pavlov *et al.* (2003) suggested (1.0-3.0) $10^{-4}$ vmr $CH_4$ and Selsis (2000) implied an upper limit of 1 $10^{-3}$ vmr. Such high values arose from low hydroxyl concentrations, an important sink for $CH_4$ during the Proterozoic. OH was low because it was consumed by high levels of Proterozoic CO and $H_2$ (Brown, 1999).

High CO levels from $CO_2$ photolysis could build up during the Archaean. Kasting and Catling (2003) suggested 5.5 $10^{-3}$ vmr. This would also favour the smog mechanism – here, analagous series of $O_3$-forming reactions occur but with CO replacing $CH_4$. Proterozoic CO, however, is not well-determined. Part of the uncertainty arises from opposing effects, as discussed in Pavlov and Kasting (2002). On the one hand during the *early* Proterozoic, $O_2$ and $O_3$ were very low which meant that high levels of UV flux reached the surface and were able to photolyse water vapour to produce OH. As $O_2$ increased further this effect weakened leading to decreased OH (hence increased CO, $H_2$). On the other hand during the *mid to late* Proterozoic (see Figure 1), there exists an opposing mechanism, which increases OH. So rising $O_2$ (hence $O_3$) lead to increased OH (hence decreased CO, $H_2$) via $O_3$+hν $O_2$+O($^1$D), then O($^1$D) +$H_2O$ 2OH. In our runs, we adopted a factor of 10 *decrease* for Proterozoic CO compared with Archaean values, corresponding to the mechanism operating in the mid to late Proterozoic. Model tests suggest however, that the uncertainty in this compound does not greatly affect the conclusions - the $O_3$ column was far more sensitive to changes in $CH_4$ and $NO_x$.

$H_2O$ indirectly affects the smog mechanism, because it photolyses to produce $HO_x$ (=OH+$HO_2$), which removes $NO_x$ (=NO+$NO_2$) into inactive, reservoir forms e.g. OH+$NO_2$+M $HNO_3$ (nitric acid)+M, or via OH+NO+M HONO (nitrous acid)+M. 'M' represents any available species (typically $N_2$) required to carry away excess energy from the reaction which would otherwise cause the products immediately to fall apart. Surface $H_2O$ in our model was set to vary sinusoidally in all runs from (1.0-1.3) $10^{-2}$ vmr from midnight to midday respectively, consistent with Kasting (1997) and Kasting and Catling (2003). $H_2$ indirectly affects the smog mechanism by shifting the OH/$HO_2$ ratio to favour $HO_2$. Kasting and Catling (2003) suggested Archaean $H_2$ of 1 $10^{-3}$ vmr. This value was derived by balancing volcanic emissions with loss rates via escape to space. Tian *et al.* (2005) recently argued that loss to space may have occurred up to 100 times slower than previously thought, which would imply a much higher value for Archaean $H_2$ of up to 0.3 vmr. In our runs we adopted a factor of 10 decrease for Proterozoic $H_2$ compared with the Archaean value from Kasting and Catling (2003).

$O_2$ is required by the smog mechanism to form $O_3$ via $O+O_2+M$  $O_3+M$. Kasting and Catling (2003) provide a review on the rise of atmospheric $O_2$. Sulphur isotope data (Farquhar *et al*., 2001; Pavlov and Kasting, 2002) and trace sulphate data (Hurtgen *et al*., 2002) suggests that $O_2$ rose to at least 21 ppmv 2.3 billion years (Ga) ago, termed the "Great Oxidation Event" (GOE) and believed to be associated with cyanobacteria. The GOE may have been favoured by high methane levels because methane can diffuse into the upper atmosphere where it photolyses and subsequently loses some of its hydrogen, which is a strong sink for $O_2$, to space (Kerr, 2005). $O_2$ then rose again, i.e. the "Second Oxidation Event" (SOE) between 0.6 to 0.8 Ga ago. Many scientists contend that the SOE was related to an increase in the removal rate of organic carbon rapidly consuming $O_2$ into sediment but the exact mechanism is not clear (Kerr, 2005). Lenton *et al*. (2004) suggested that the SOE  was linked with increased weathering of rocks which led to more phosphorous  (a nutrient) being released into the seas, favouring more $O_2$ production. The increased weathering may have been due to the formation of a supercontinent or/and  the arrival of lichens on land (Kerr, 2005). Typical "background" $O_2$ Archaean values are $10^{-14}$ to $10^{-18}$ vmr usually quoted based on model output (Kasting, 1993) with Proterozoic values typically varying between 0.1 up to 10% present day $O_2$.

$NO_x$ is required to catalyse the $O_3$ smog mechanism. In today's atmosphere, about 50 $10^{12}$ g/year (or 50Tg/year) of nitrogen are emitted into the atmosphere in the form of $NO_x$ (Houghton *et al*., *Intergovernmental Panel on Climate Change (IPCC) Third Assessment Report (TAR)*, 2001). 60% comes from industry, most of the remainder comes from soils, lightning, and biomass burning. Typical modern day $NO_x$ concentrations vary from about 10 pptv in very clean air, up to 100 ppbv in very polluted air. In the abiotic atmosphere $NO_x$ could have built up to 1ppmv from natural sources (Selsis *et al*., 2002b) because OH, an important $NO_x$ sink, was low. Navarro-González *et al.* (2001) suggested a $NO_x$ source of 1-10 Tg N/year in the abiotic atmosphere based on simulated lightning-discharge experiments. Cosmic Rays (CRs) associated with an active sun may lead to $N_2$ dissociation hence $NO_x$ formation in the middle atmosphere at high latitudes (Callis *et al*., 2001, Randall *et* *al*., 2001). Recent model studies (e.g. Langematz *et al*., 2005) suggest a potentially large effect i.e. $NO_x$ changes in excess of 10 ppbv, despite large uncertainties. In the early Earth, this effect may have been even larger because the high energy output of the Sun (X-ray, solar wind density) was up to 100-1000 times higher than present day (Guinan and Ribas, 2002; Ribas *et al*., 2005). Smith *et al.* (2004) investigated the effects of ionising radiation (e.g. stellar flares, galactic cosmic rays, energetic particles) in terrestrial-like exoplanets from a radiative standpoint, but without considering $NO_x$ formation mechanisms.

To decide the level of $NO_x$ to be set in our study, Table 1 gives an overview of $NO_x$ sources for the early Earth atmosphere and the pre-industrial year 1850 atmosphere. Perhaps most interesting is the Cosmic Ray mechanism. Uncertainties here are large, but the Cosmic Ray source strength is also potentially very large, because EUV enhancement factors for the early Earth of 100 have been suggested (Guinan and Ribas, 2002). In today's atmosphere, Cosmic Rays produce $NO_x$ in-situ in the middle atmosphere polewards of 60 degrees. Model studies (Rozanov *et al*., 2004; Langematz *et al*., 2005) suggest that a significant $NO_x$ signal can then spread to lower altitudes and latitudes. For the early Earth, we assumed a $NO_x$ soil source varying from present day up to double present day values. To estimate the ambient $NO_x$ concentrations from the sources in Table 1, firstly we sum the various sources in Table 1 ignoring the uncertain Cosmic Ray source:

Sum Pre-Industrial $NO_x$ sources in Table 1 = (6.5-17.5) Tg N /year
Sum early Earth $NO_x$ sources in Table 1 = (7.5-31.0) Tg N /year
$NO_x$ factor, (early Earth/ Pre-Industrial) = (7.5-31.0)/(6.5 -17.5) = (0.43–4.77)

The Pre-Industrial General Circulation Model (GCM) study of Grenfell *et al.* (2001) without Cosmic Rays suggested an ambient Pre-Industrial $NO_x$ concentration of 0.32 ppbv. Therefore we assume:

early Earth $NO_x$ concentration= Pre-Industrial $NO_x$ concentration *$NO_x$ factor
= 0.32 ppbv * (0.43-4.77)

$$= (0.14 – 1.53) \text{ ppbv}$$

Based on the above, we performed sensitivity runs with $NO_x$ set to 0.14 and 1.53 ppbv.

The above analysis does not consider changes in OH between the present day and the early Earth. However early Earth OH is uncertain and subject to opposing mechanism as discussed earlier.

We also performed runs with $NO_x$ set to 20 ppbv. This represents an upper limit for the Cosmic Ray mechanism and assumes that $NO_x$ produced by this mechanism in the mid to upper atmosphere at high latitudes is able to penetrate into the troposphere. The extent to which this occurs is currently a focus of research for the Earth modelling community e.g. (Sinnhuber *et al.*, 2003) suggested 10-20 ppbv $NO_x$ could reach at least down to 20 km whereas (Quack *et al.*, 2001) suggested that such values remained above 40 km.

An interesting point is, $NO_x$ and therefore also ozone on the early Earth would clearly have been enhanced close to its sources e.g. massive volcanoes which generated their own lightning. So, even if $NO_x$ could not build up globally because it was destroyed by high OH, nevertheless individual "islands" of $NO_x$ with associated ozone could have survived close to $NO_x$ sources. We envisage that these would have offered at least isolated pockets of shielding from UV hence encouraged the establishment of life in a heterogeneous manner. We quantify these ideas in the results section.

### 2.2 Computational details

Our approach here is somewhat different to other modelling studies of the early Earth (e.g. Kasting, 1993; Segura *et al.*, 2003). Those studies typically employed column models integrated until specified emission rates of source gases reach equilibrium with their sinks. Photolysis rates in those studies were diurnally averaged. Our model takes the approach of a box model, as used in modern day atmospheric chemistry measurement campaigns. It solves a chemical reaction network for a particular location and time with assumed source gas concentrations, temperature, humidity, pressure etc. Then, important reactive intermediates such as $O_3$ and OH are calculated including diurnal variations, In this work, we vary source gases: $O_2$, $H_2$, CO, $CH_4$ and $NO_2$ within reasonable bounds of uncertainty for the early Earth, then use our model to calculate the reactive intermediates. Advantages of the boxmodel approach are:

(1) The accurately validated output compares well with present day observations.
(2) Attributing chemical responses between runs is relatively straightforward.
(3) The model requires a short integration time for the reactive intermediates to adjust to the specified source gas concentrations, hence:
(4) A large number of sensitivity runs may be performed.

These calculations can also act as a basis for a column model developed in the future.

We used the FACSIMILE commercial integrating package developed by the Materials and Chemical Process Assessment (MCPA). Chemical reactions were treated as a system of non-linear, time-dependent differential equations and are solved using the Gear-method (Gear *et al*., 1985, and references within). Our straightforward tropospheric chemical scheme featured inorganic $HO_x$, $NO_x$ and $O_x$ reactions and comprised 55 photochemical reactions for 25 species, 10 of which were photolysis reactions, and including $CH_4$ oxidation. The basic scheme is available at http://mcm.leeds.ac.uk/MCM. The full set of equations solved here is given in Appendix I.

We ran for perpetual[2] June conditions at a latitude of 53$^o$N, corresponding to the atmospheric measurement station at Weybourne in England. This latitude was originally chosen for present Earth model validation. In the absence of Proterozoic data, temperature in the model varied sinusoidally from 290 K to 298 K from midnight to midday respectively and relative humidity varied from 44% to 90% from midnight to midday respectively.

[2] Solar zenith and local hour angles are calculated based on mean June conditions, hence the length and amplitude of the diurnal cycle in flux is constant from day to day.

Photolysis rates (J) were of the form: ( J = K cos(sza) L exp( -M sec(sza) ) ) where K, L, M are molecule-dependent constants, sza=solar zenith angle. These expressions were derived from previous runs of a GCM, as described in Hough (1988). The model included a parameterisation to calculate mixing height (H) using a time-dependant triangular spike function which varied from 300m at midnight to 1300m at midday. The variation reflects heating of the surface during the daytime leading to convective currents pushing the mixing height upwards. Calculating the mixing height was important because it determined the rate of loss of species to the ground. This was parameterised via so-called deposition velocities (D (cm $s^{1}$)). Physically these represent the "stickiness" of a species i.e. how quickly it may travel to the surface and be permanently removed there. Numerically, rate of loss via deposition in the model, $k_{dep}$ = (D/H) $s^{-1}$, where D($HNO_3$)=2.0 cm $s^{-1}$, D($NO_2$)=0.15 cm $s^{-1}$, D($O_3$)=0.5 cm $s^{-1}$, D($H_2O_2$)=1.1 cm $s^{-1}$,
D(HCHO) (formaldehyde)= 0.33 cm $s^{-1}$, D($CH_3NO_3$)=1.1 cm $s^{-1}$,
D($CH_3OOH$) = 0.55 cm $s^{-1}$.
Photolysis rates for the early Earth ($J_{early\ Earth}$) were calculated for a particular species via:

$$J_{early\ Earth} = J \ ?F \ ?C$$

J represents present day photolysis rates, F is the faint Sun factor and C is a factor depending on the overhead ozone column. Numerically, F = $(1+0.4(1-t/t_o))^{-1}$ where $t_o$ is the present age of the Earth (=4.6Ga), and t is the age of the early Earth (Ga). For our runs we took t to be 3.9 Ga i.e. around the time of the second oxidation event (see Figure 1). This resulted in an F value of 0.943. The column factor, C, represents increases in ground UV due to a weaker stratospheric $O_3$ column for the early Earth. We derived C values from Segura *et al*. (2003) who performed calculations with, for example, 1% and 10% $O_2$ using a column model. Those authors split C-factor contributions into their UV-A, UV-B and UV-C components (refer to their Table 2). In order to be consistent with the Segura C-values, we also split photolysis processes in our model into contributions from UV-A, UV-B, and UV-C, as shown in Table 2.

In total thirty six sensitivity runs were performed, summarised in Table 3. The fixed species in the model were varied from run to run to represent uncertainties in their concentrations (see section 2.1). The response time is relatively short because the long-term source gases in the model are specified by the user. The model runs reached equilibrium after about 10-20 days. However, the model sometimes experienced numerical problems whereby the integrator failed to converge. This usually happened at dawn or/and dusk when chemical concentrations changed rapidly and drastically. The problem affected the 1% PAL $O_2$ runs since these represented the weakest ozone cover, therefore featured the highest surface photolysis fluxes (see Figures 2a, 3a). The convergence problem also occurred at night. This could mean that the constraints we imposed i.e. constant $NO_x$, $CH_4$, CO, etc. are unrealistic or it could indicate some missing nighttime chemistry of the early Earth. Only calculations which did reach equilibrium were included in Figures 2 and 3.

## 3. Results

### 3.1 Ozone

Figure 2 shows surface concentrations of ozone for (a) 1% and (b) 10% PAL $O_2$. Some midnight data are missing because the model sometimes featured convergence problems during the night, as already discussed. The values shown are all much enhanced compared with our pre-industrial control run, which featured midday $[O_3]=2.8\ 10^{-9}$ mixing ratio. Results in Figures 2a, 2b may be broadly split into three categories, namely low, medium and high ozone which corresponded to low (=0.14 ppbv), medium (=1.53 ppbv) and high (=20 ppbv) $NO_x$. The results suggest for the low and medium $NO_x$ cases, that ozone was controlled mainly by changes in $NO_x$ rather than changes in $CH_4$, $H_2$, or CO. This is analagous to the present-day situation in many unpolluted regions of the Earth, which are said to be under “$NO_x$ control” with regard to ozone smog production (e.g. Shindell *et al.*, 2001). For the high $NO_x$ runs, however, ozone production appears to get "saturated" with respect to $NO_x$ and responds instead to changes in $H_2$, CO and $CH_4$. This is analagous to the situation today in some large cities which are said to be under "hydrocarbon control" as regards ozone smog production. For the high $NO_x$ runs, increasing $H_2$, CO and $CH_4$ one by one by the amounts shown in Table 3 lead to an increase in midday $O_3$ from about (2.4-3.6) ppmv for the 1% PAL $O_2$ runs (Figure 2a) and from (0.02-0.36) ppmv at midday for the 10% PAL $O_2$ runs (Figure 2b).

Appendix II shows how we calculated the ozone column in DU using run 9 as an example (i.e having medium levels of $NO_x$ and VOCs). We assume a 12 km troposphere with constant ozone values as calculated by our box-model for the surface. The calculation implies 564.3 DU for the troposphere of the Early Earth i.e. well in excess of today's total (tropospheric+stratospheric) ozone column of (300-350) DU.

**3.2 Hydroxyl (OH)**

Figures 3a, 3b are the same as Figures 2a, 2b but show hydroxyl. Our pre-industrial control run featured midday [OH] = 3.34 $10^6$ molecules/$cm^3$ i.e. the 1% PAL $O_2$ runs were enhanced by approximately a factor of 1000 compared with the control run but OH for the 10% PAL of $O_2$ runs were suppressed. We interpret this as follows: on the one hand, higher fluxes favour more OH but on the other hand, less $O_2$ favours less OH compared with the control run. For the 1% PAL $O_2$ runs, it would seem that high fluxes lead to higher OH but for the 10% PAL $O_2$ runs, less $O_2$ leads overall to less OH compared with the control run.

OH concentrations for a particular time of day varied hardly at all in the model after a relaxation period of about 5 days. In Figures 3a, 3b, increasing CO, $CH_4$ or $H_2$ leads to a lowering in OH by a factor of 3-7. This is the expected result, since these compounds react directly with OH. Also, we see in both Figures a slight upward trend with increasing run number at midday i.e. with increasing $NO_x$. This effect has been studied (Prather *et al.*, 2003) and arises via $HO_2+NO \rightarrow OH+NO_2$ i.e. as $NO_x$ increases, $HO_2$ repartitions leading to an increase in OH. The upward trend is less clear for the midnight values especially for the high $NO_x$ case, which could point to other processes playing a role e.g. at high $NO_x$, $NO_2$ becomes a *sink* for OH via $OH+NO_2+M \rightarrow HNO_3+M$ although the lack of data at midnight in Figure 2a makes interpretation difficult.

Figures 4a, 4b show the rates of the major sinks $CH_4$+OH, $NO_2$+OH, $H_2$+OH, $NO_2$+OH and the major source for OH ($O(^1D)+H_2O \rightarrow 2OH$) for the control run and for "typical" early Earth run[3]. For the control run, the source $O(^1D)+H_2O$ proceeds faster than the sinks, which all proceed at an approximately comparable rate. For the early Earth run, the CO, $CH_4$ and $H_2$ sinks are all important – clearly any of these three could be the major remover of OH within their reported uncertainty range.

## 4. Discussion

### 4.1 Ozone

The 1% PAL $O_2$ runs in Figure 2a featured much higher daytime ozone levels (0.1 to 3.6 ppmv) compared with the 10% PAL $O_2$ runs in Figure 2b (0.004 to 0.36 ppmv). This was associated with enhanced photolytic fluxes in the 1% PAL $O_2$ runs due to a weaker assumed overhead ozone column, which was coded into the model via the "column factors" taken from the Segura study. The fact that stronger photolysis fluxes stimulate the ozone smog mechanism is widely accepted – ozone smog is a much greater problem in sunny cities such as Athens and Los Angeles, with increased flux leading to more ozone smog as already discussed.

Midnight ozone for the 1% PAL $O_2$ runs is higher than its midday counterpart (Figure 2a) for low $NO_x$ (runs 1 – 6). In today's atmosphere, important mechanisms affecting nocturnal $O_3$ near the surface are: (1) dry deposition of $O_3$ to the ground and (2) chemical removal of $O_3$ via $NO+O_3 \rightarrow NO_2+O_2$. Regarding (1), the rate of dry deposition of ozone equals $k_{dep}(O_3)$, as already discussed in section 2.2. The 1% PAL $O_2$ runs featured higher $O_3$ hence also *faster loss* to the ground compared with the 10% PAL $O_2$ runs so clearly mechanism (1) cannot lead to higher midnight $O_3$ values in the 1% PAL $O_2$ case. Further checking revealed that mechanism (2) also could not explain the high midnight $O_3$ for the 1% PAL $O_2$ runs either. In fact, the 1% PAL $O_2$ runs featured higher midnight NO than the 10% PAL $O_2$ runs which would imply less $O_3$. The higher NO was related to lower midnight OH for the 1% PAL $O_2$ case, hence slower removal of NO into its sinks. The lower midnight OH was in turn related to the stronger daytime oxidation of methane in the case of the high flux 1% PAL $O_2$ run leading to higher concentrations of methane oxidation products such as HCHO, $CH_3OOH$, which perturbed the night chemistry and acted as sinks for OH.

[3] In this case run 19, in which $(O_2)$=10% PAL and the long-lived source gases are in the mid-range of their reported values.

To understand why midnight $O_3$ was higher in the 1% PAL $O_2$ runs 1 - 6, it was helpful to consider ozone concentrations over the diurnal cycle. In both the 1% PAL $O_2$ and 10% PAL $O_2$ cases there occurred an $O_3$ peak in the afternoon. This arose because typical response timescales of the smog cycle were a couple of hours, which led to $O_3$ levels lagging the midday peak in solar intensity. However, the 1% PAL $O_2$ runs, feature stronger daytime fluxes. This led to higher ozone at dusk, which persisted until midnight, hence the high midnight values of run 1-6.

Increasing $NO_x$ from the medium to the high case (Figures 2a, b) led to a rise in ozone by a factor of 2-3. Why? At high $NO_x$, the smog mechanism appears to get saturated and an opposing mechanism can play a role in which ozone is directly removed via NO. This behaviour is sometimes also seen in the centres of large cities, where NO is high due to dense traffic but where ozone levels can be actually lower than in the suburbs. This low ozone effect is suggested by run 33.

Changing CO, $CH_4$, $H_2$ had a much smaller effect on ozone compared with the $NO_x$ changes. For the low $NO_x$ case, changing these three compounds (Figures 2a, 2b) had only a very small effect on ozone.

For the medium $NO_x$ case CO and $CH_4$ changes typically resulted in small ozone volume mixing ratio changes of (2-5) $10^{-8}$. Increasing $H_2$ by a factor of 10 (runs 7-8, 25-26) led to an ozone increase of 0.05 $10^{-6}$. Why? In the model, there is only one chemical reaction involving $H_2$ namely: $OH+H_2+O_2$-->$HO_2+H_2O$. So, $H_2$ can affect the $HO_x$ (hence $NO_x$) partitioning affect ozone, although the effect is small. The important point is, the overall effect of increasing $H_2$ leads to a small ozone increase in the model.

For the high $NO_x$ runs, changing CO, $CH_4$ and $H_2$ led to significant changes in ozone (Table 4, Figures 2a, 2b). For the 1% PAL $O_2$ runs (Figure 2a) changing these three compounds individually led to ozone varying between 2.2 $10^{6}$ and 3.6 $10^{-6}$. More $CH_4$ and CO we simply interpret as leading to a faster smog mechanism hence more ozone. For the high $NO_x$ runs the $H_2$ effect already discussed for the medium case runs is now stronger. Here, increasing $H_2$ from modern-day values to Proterozoic values leads to large increases in ozone e.g. from about 0.030 ppm in run 33 to 0.36 ppm in run 34 (Figure 2a,b).

**4.2 Hydroxyl**

The OH values in Figures 3a, 3b can be used to constrain a $NO_x$ lifetime in the early Earth, hence estimate the size of  the proposed UV "protective pockets" in the vicinity of natural $NO_x$ sources. Assume for the 1% PAL $O_2$ runs,  early Earth OH is 1000 times higher than today's levels. This is implied by Fig. 3a which shows peak OH of about 1 $10^{9}$ molecules/cm whereas typical modern day peak 1 $10^{6}$ i.e. 1000 times less. In the modern atmosphere, OH is converted into nitric acid on a timescale of about two weeks. So, for the early Earth, 1000 times more OH implies a lifetime of (2 weeks/1000) = about 20 minutes. A typical air parcel in the boundary layer may travel up to 1km during this time period. We do not consider what occurs before 1% PAL  $O_2$, when the fluxes, hence OH, were very high and therefore the UV "protective pockets" must have been very small. We note in passing however, that volcanoes emit dust and ash which also scatter UV hence protect the surface – this  may also have played a role, especially at the very start of the Proterozoic.

## 5. Conclusions

The results of this work imply that the $O_3$ smog mechanism has the potential to shield a significant amount of harmful UV from the surface during the Proterozoic era. Due to positive feedback - in the sense that life produces some $O_2$, which then forms some $O_3$ via the smog mechanism under whose shield more life can propagate, and so on - the mechanism is therefore self-reinforcing.

One may consider on the other hand, that high levels of tropospheric $O_3$ protect life from UV but are also toxic to many of today's plants and animals. However, its toxic effect upon early procaryote cells is not well-determined.

Many works e.g. von Bloh *et al.* (2003) have investigated the "Pre-Cambrian explosion" characterised by rapid increase in biomass, occuring at the end of the Proterozoic. Our results suggest a positive feedback, in which an initial small rise in $O_2$ associated with life, leads to tropospheric $O_3$ production which favours the establishment of more life and so on. We are planning to extend this study using a column model of the early Earth. In a column model, $O_3$ formed in the upper troposphere would protect the surface from UV hence slow the smog mechanism. This could lead to lower $O_3$ colums than the values shown in this sensitivity study.

The results obtained are sensitive to the amount of $NO_x$ assumed, the sources of which are rather uncertain. Our medium value (=1.53 ppbv $NO_x$) runs with 1% PAL $O_2$ produce around 1.0 $10^{-6}$ surface ozone mixing ratio. Assuming this concentration exists throughout the troposphere implies a column value of around double that of the present day (see Appendix II).

Clearly, this result should be checked with a column model, to include the effects of self-healing i.e. ozone increases on upper levels block the passage of UV to lower levels. $NO_x$ values in Table 1 represent global mean averages. However, ambient concentrations of local $NO_x$ may have built up to much higher concentrations in the vicinity of "hotspots" e.g. near volcanoes in the early Earth, so the smog mechanism may have evolved in particular locations without requiring large global $NO_x$ averages.

Regarding the anticipated search programs for Earthlike planets, our results point to the possibility of a strong tropospheric $O_3$ signal in early developing atmospheres which may be indicated in observations by pressure-broadening effects of spectral lines. Given the large number of young stars in our solar neighborhood, effort to understand the atmospheric composition of the early Earth is an important focus for the coming decade.

**Acknowledgments**

We thank the authors of the Master Chemical Mechanism, Leeds University, UK for providing the original chemical scheme.

**Appendix I – Overview of the Chemical Scheme**

*30 thermal inorganic gas-phase reactions:*

$O+O_2+N_2$-->$O_3+N_2$

$O+O_2+O_2$-->$O_3+O_2$

$O+NO$-->$NO_2$

$O+NO_2$-->$NO+O_2$

$O+NO_2+M$-->$NO_3+M$

$O(^1D)+N_2$-->$O+N_2$

$O(^1D)+O_2$-->$O+O_2$

$NO+O_3$-->$NO_2+O_2$

$NO_2+O_3$-->$NO_3+O_2$

$NO+NO+O_2$-->$NO_2+NO_2$

$NO+NO_3$-->$NO_2+NO_2$

$NO_2+NO_3$-->$NO+NO_2+O_2$

$NO_2+NO_3+M$-->$N_2O_5+M$

$O(^1D)+H_2O$-->$OH+OH$

$OH+O_3$-->$HO_2+O_2$

$OH+H_2+O_2$-->$HO_2+H_2O$

$OH+CO+O_2$-->$HO_2+CO_2$

$OH+H_2O_2$-->$HO_2+H_2O$

$HO_2+O_3$-->$OH+2O_2$

$HO_2+HO_2+O_2$-->$H_2O_2+O_2$

$HO_2+HO_2+N_2$-->$H_2O_2+N_2$

$OH+NO+M$-->$HONO+M$

$OH+NO_2+M$-->$HNO_3+M$

$OH+NO_3$-->$HO_2+NO_2$

$HO_2+NO$-->$OH+NO_2$

$HO_2+NO_2+M$-->$HO_2NO_2+M$

$OH+HO_2NO_2$-->$NO_2+H_2O+O_2$

$HO_2+NO_3$-->$OH+NO_2+O_2$

$OH+HONO$-->$NO_2+H_2O$

$OH+HNO_3$-->$NO_3+H_2O$

*10 inorganic photolysis reactions*

$O_3$ -->O($^1$D)+$O_2$

$O_3$-->O+$O_2$

$H_2O_2$-->OH+OH

$NO_2$-->NO+O

$NO_3$-->NO+$O_2$

$NO_3$-->$NO_2$+O

HONO-->OH+NO

$HNO_3$-->OH+$NO_2$

$CH_3NO_3$-->$CH_3O$+$NO_2$

$CH_3OOH$-->$CH_3O$+OH

*15 methane oxidation reactions*

OH+$CH_4$+$O_2$-->$CH_3O_2$+$H_2O$

$CH_3O_2$+NO-->$CH_3O$+$NO_2$

$CH_3O_2$+NO-->$CH_3NO_3$

$CH_3O$+$O_2$-->HCHO+$HO_2$

$CH_3O_2$+$NO_2$-->$CH_3O_2NO_2$

$CH_3O_2NO_2$-->$CH_3O_2$+$NO_2$

$CH_3O_2$+$NO_3$-->$CH_3O$+$NO_2$+$O_2$

$CH_3O_2$+$HO_2$-->$CH_3OOH$+$O_2$

$CH_3O_2$-->$CH_3O$+O

$CH_3O_2$-->HCHO+OH

OH+$CH_3NO_3$-->HCHO+$NO_2$+$H_2O$

$CH_3NO_3$-->$CH_3O$+$NO_2$

OH+$CH_3OOH$-->$CH_3O_2$+$H_2O$

OH+$CH_3OOH$-->HCHO+OH+$H_2O$

$CH_3OOH$-->$CH_3O$+OH

## Appendix II - Calculation of column $O_3$

Below is an example of the column calculation for run 9 ($O_3$=849.1ppbv). This run represents mean Proterozoic conditions. It features 1% $O_2$, medium $NO_x$ and $CH_4$, CO, $H_2$ in the mid-range of their reported values. After the table follows a calculation which illustrates how, for example, we arrived at the value 80.5 DU in the 0-1km interval. Other values in the Table were obtained in a comparable manner.

| Height (km) | Pressure(mb) | T(K) | M($10^{19}$ molec/cm$^3$) | Column (DU) |
|---|---|---|---|---|
| 0 | 1013.3 | 280.0 | 2.55 | 80.5 |
| 1 | 877.2 | 273.5 | 2.26 | 71.4 |
| 2 | 759.6 | 267.0 | 2.01 | 63.3 |
| 3 | 657.8 | 260.5 | 1.78 | 56.2 |
| 4 | 569.7 | 254.0 | 1.58 | 49.9 |
| 5 | 493.3 | 247.5 | 1.40 | 44.3 |
| 6 | 427.2 | 241.0 | 1.25 | 39.4 |
| 7 | 369.9 | 234.5 | 1.11 | 35.1 |
| 8 | 320.3 | 228.0 | 0.99 | 31.3 |
| 9 | 277.4 | 221.5 | 0.88 | 27.9 |
| 10 | 240.2 | 215.0 | 0.79 | 24.9 |
| 11 | 208.0 | 215.0 | 0.68 | 21.5 |
| 12 | 180.1 | 215.0 | 0.59 | 18.6 |
| | | | | ----------- |
| | | | **Total Column =** | **564.3 DU** |

Pressure (mb), P = $1013.3/10^{(z/16)}$ , where z= height (km).

T (K) was taken from US standard atmosphere (1976).

Total number density (M) = $M_{ground}*(P/P_{ground})*(T_{ground}/T)$.

Assume 1 Dobson Unit (DU[4]) = 2.69 $10^{16}$ molecules/cm$^2$.

Column (DU) = M×$O_3$ vmr×100000 (1km thickness in cm) / DU factor

e.g. for z=0km: 2.55 $10^{19}$×849.1 $10^{-9}$×100000 / (2.69 $10^{16}$) DU = 80.5 DU

[4]DU represents 0.01 mm $O_3$ column at 0° Celsius and 1 atm pressure. Typical values are 300-350 DU, i.e. 3-3.5 mm $O_3$ column at standard temperature and pressure.

**References**

Bada, J. L (2004) How life began on earth: a status report, *Earth Plan. Sci. Lett.* 226, 1-15, 2004.

Beichman, C.A, N. J. Woolf, and C. A. Lindensmith (eds.) (1999) The Terrestrial Planet Finder (TPF): a NASA origins program to search for habitable planets (JPL Publication 99-003 available at tpf.jpl.nasa.gov).

von Bloh, W., C. Bounama, S. Franck (2003) Cambrian explosion triggered by geosphere-biosphere feedbacks, *Geophys. Res. Lett.* 30.

Brown, L. L. (1999) Numerical models of reducing primitive atmospheres on Earth and Mars, Ph.D thesis, Penn State University.

Callis, L.B., M. Natarajan, and J.D. Lambeth (2001) Solar-atmospheric coupling by electrons (SOLACE)3. Comparisons and observations, 1979-1997, issues and implications, *J. Geophys. Res.* 106, 7523-7539.

Cleaves, H. J., and S. L. Miller (1998) Oceanic protection of prebiotic organic compounds from UV radiation, *Proc. Natl. Acad. Sci.*, USA, 95, 7260-7263.

Demerjian, K. L., J.A. Kerr and J.G. Calvert (1974) The mechanism of photochemical smog, *Advan. Environ. Sci. Technol.*, Vol. 4, 1-262, Wiley Interscience, New York, NY.

DeMore, W. B., S. P. Sander, D. M. Golden, R. F. Hampson, M. J. Kurylo, C. J. Howard, A. R. Ravishankara, C. E. Kolb, and M. J. Molina (1994) Chemical kinetics and photochemical data for use in stratospheric modeling, Jet Propulsion Laboratory, Pasadena, California.

Farquhar, J., J. Savarino, S. Airieau, and M. H. Thiemens (2001) Observation of wavelength sensitive, mass-independent sulfur isotope effects during $SO_2$ photolysis: application to the early atmosphere, *J. Geophys. Res.* 106, 32,829 -32,839.

Gear, C. W., G. K. Gupta, and B. Leimkuhler (1985) Automatic Integration of Euler-Lagrange Equations with Constraints, *J. Comp. and App. Math.*, 12-13, 77-90.

Grenfell, J.L., D.T. Shindell, D. Koch, and D. Rind (2001) Chemistry-climate interactions in the Goddard Institute for Space Studies general circulation model: 2. New insights into modeling the pre-industrial atmosphere. *J. Geophys. Res.* 106, 33435-33451.

Guinan, E. F., and I. Ribas I. (2002) Our changing Sun: the role of solar nuclear evolution and magnetic activity on Earth's atmosphere and climate, in *The Evolving Sun and its Influence on Planetary Environments (B. Montesinos, A. Gimenz & E. F. Guinan, (Eds.), 269, 85–107, ASP, San Francisco*.

Haagen-Smit, A. J., C. E. Bradley, and M. M. Fox (1952) Formation of Ozone in Los Angeles Smog. Proceedings of the Second National Air Pollution Symposium, Discussions on Fundamental Chemistry and Physics of the Atmosphere.

Hough, AM (1988) The Calculation of Photolysis Rates for Use in Global TroposphericModelling. Studies, AERE Report R-13259, HM Stationery Office, London.

Houghton, J. T., Y. Ding, D. J. Griggs, M. Noguer, P. J. van der Linden, X. Dai, K. Maskell, and C. A. Johnson (eds) (2002) Intergovernmental Panel on Climate Change (IPCC), Climate Change 2001: The Scientific Basis. Contribution of Working Group 1 to the Third Assessment Report, , Cambridge University Press, United Kingdom and New York, 881pp.

Hurtgen, M. T., M. A. Arthur, N. Suits, and A. J. Kaufman (2002) The sulfur isotopic composition of neoproterozoic seawater sulfate: implications for a snowball earth, *Earth. Plan. Sci. Lett.* 203, 413-430.

Kasting, J. F. (1993) Earth's early atmosphere, *Science* 259, 920-926.

Kasting, J. F. (1997) Origins of life and evolution of the biosphere, 27, 291-307, Kluwer academic publications, The Netherlands.

Kasting, J. F., and T. M. Donahue (1980) The evolution of atmospheric ozone, *J. Geophys. Res.* 85, 3255-3263.

Kasting, J. F., and D. Catling (2003) Evolution of a habitable planet, *Ann. Reviews Ast. Astrophys.* 41, 429-463.

Kerr, R. A. (2005) The story of $O_2$, *Science* 308, 1730-1733.

Kumar, P., and D. Mohan (2002) Photochemical smog mechanism, ill-effects and control, TIDEE 1(3), 445-456.

Lammer, H., E. Chassefiere, Y. N. Kulikov, F. Leblanc, H. I. M. Lichtenegger, J. -M. Griessmeier, M. Khodachenko, D. Satm, C. Sotin, I. Ribas, F. Selsis, F. Allard, I. Mingylev, O. Mingalev, H. Rauer, J. L. Grenfell, D. Langmayr, G. Jaritz, S. Endler, G. Wuchterl, S. Barabash, H. Gunell, R. Lundin, H. K. Biernat, H. O. Rucker, F. Westall, A. Brack, S. J. Bauer, A. Hanslmeier, P. Odert, M. Leitzinger, P. Wurz, E. Lohinger, R. Dvorak, W: W. Weiss, W. von Bloh, S. Franck, T. Penz, A. Stadelmann, U. Motschann, N. K. Belisheva, A. Berces, A. Leger, C.S: Cockell, J.

Parnell, I. L. Arshukova, N. V. Erkaev, A. A. Konovalenko, C. Moutou, F. Forget, B. Erdi, A. Hatzes, E. Szuszkiewicz, and M. Fridlund (2005) Towards real comparative planetology: synergies between solar system science and the Darwin mission, manuscript submitted.

Langematz, U., J. L. Grenfell, K. Matthes, P. Mieth, M. Kunze, B. Steil, and C. Brühl (2005) Chemical Effects in 11-year solar cycle simulations with the Freie Universität Berlin Climate Middle Atmosphere Model with online chemistry (FUB CMAM CHEM), *Gephys. Res. Lett.* 32, L13803.

Léger A., J. M. Mariotti, B. Mennesson, M. Ollivier, J. L., Puget, D., Rouan, and J. Schneider (1996) Could we Search for Primitive Life on Extrasolar Planets in the Near Future? The DARWIN Project, *J. Astrophys. Spa. Sci.*, 241 (1), 135-146.

Léger, A. (2000) Strategies for remote detection of life – DARWIN-IRSI and TPF Missions, *Adv. Space Res.* 25, 2209-2223.

Lenton, T. M., H. J. Schellnhuber, and E. Szathmáry (2004) Climbing the co-evolution ladder, *Nature* 431, 913.

Levine, J. S., P. B. Hays, and J. C. G. Walker (1979) The evolution and variability of atmospheric ozone over geological time, *Icarus* 39, 2, 295-309.

Navarro-González, R., C. P. McKay, D. N. Mvondo (2001) Possible nitrogen crisis for Archaean life due to reduced nitrogen fixation by lightning, *Nature* 412, 61-64.

Mvondo , D. N., R. Navarro-González, C. P. McKay, P. Coll, F. Raulin (2001) Production of nitrogen oxides by lightning and coronae discharge in simulated early Earth, Venus and Mars enviroments, *Adv. Space. Res.*, 27, 2, 217-223.

Pavlov, A. A., J. F. Kasting, and L. L. Brown (2000) Greenhouse warming by $CH_4$ in the atmosphere of early Earth, *J. Geophys. Res.* 105, 11981-11990.

Pavlov, A. A., L. L. Brown, and J. F. Kasting (2001) UV shielding of $NH_3$ and $O_2$ by organic hazes in the Archaean atmosphere, *J. Geophys. Res.* 106, 23,267-23,288.

Pavlov, A.A., and J. F. Kasting (2002) Mass-independent fractionation of sulfur isotopes in Archaean sediments: strong evidence for an anoxic Archaean atmosphere, *Astrobiology*, 2, 1 27-41.

Pavlov, A. A., M. T. Hurtgen, J. F. Kasting, M. A. Arthur (2003) Methane-rich proterozoic atmosphere? *Geology*, 31 (1), 87-90.

Prather, M., M. Gauss, T. Berntsen, I. Isaksen, J. Sundet, I. Bey, G. Brasseur, F. Dentener, R. Derwent, D. Stevenson, L. Grenfell, D. Hauglustaine, L. Horowitz, D. Jacob, L. Mickley, M. Lawrence, R. von Kuhlmann, J.-F. Muller, G. Pitari, H.

Rogers, M. Johnson, J. Pyle, K. Law, M. van Weele, and O. Wild (2003) Fresh air in the 21st Century?, *Geophys. Res. Lett.* 30, 1100.

Price, C. and D. Rind (1994) Modelling global lightning distributions in a general circulation model, *M. wae. Rev*, 122, 1930-1939.

Quack, M., M. B. Kallenrode, M. von König, K. Kümzi, J. Burrows, B. Heber, and E. Wolff (2001) Ground level events and consequences for stratospheric chemistry, Proceedings of the Copernicus Gesellschaft.

Randall, C.E., D.E. Siskind, and R.M. Bevilacqua (2001) Stratospheric $NO_x$ enhancements in the southern hemisphere vortex in winter/spring of 2000, *Geophys. Res. Lett.* 28, 2385-2388.

Ribas, I., E. F. Guinan, M. Güdel, and M. Audard (2005) Evolution of the solar activity over time and effects on planetary atmospheres: I. High-energy irradiations (1-1700A), *Astrophys. J.* 622, 680-694.

Rozanov, E. V., M. E. Schlesinger, T. A. Egorova, B. Li, N. Andronova, and V. A. Zubov (2004) Atmospheric Response to the Observed Increase of Solar UV Radiation from Solar Minimum to Solar Maximum Simulated by the UIUC Climate-Chemistry Model. *J. Geophys. Res.* 109, D01110.

Rye, R., P. H. Kuo, and H. D. Holland, Atmospheric carbon dioxide concentrations before 2.2 billion years ago, *Nature* 378, 603-605, 1995.

Sagan, C., and G. Mullen (1972) Earth and Mars: Evolution of atmospheres and surface temperatures, *Science* 177, 52-56.

Sagan, C., and Chyba, C. (1997) The early faint sun paradox: organic shielding of ultraviolet-labile greenhouse gases, *Science* 276, 1217-1220.

Schöneberg, R., B. S. Kamber, K. D. Collerson, S. Moorbath (2002) Tungston isotope evidence from ~3.8Gyr metamorphised sediments for early meteorite bombardment of the Earth, *Nature* 418, 403-405.

Schopf, J. W. (1983) Earth's Earliest biosphere: its origin and evolution, Princeton University Press, Princeton, New Jersey.

Segura, A., K. Krelove, J. F. Kasting, D. Sommerlatt, V. Meadows, D. Crisp, M. Cohen, and E. Mlawer (2003) Ozone concentrations and ultraviolet fluxes on earth-like planets around other stars, *Astrobiol.*, 3, 689-708.

Selsis, F. (2000) PhD Thesis, Chapter 5, Universite de Boredeaux, available at: alienor.observ.u-bordeaux.fr/pub/selsis/selsis.pdf.

Selsis, F., D. Despois, and J. -P. Parisot (2002a) Signature of life on exoplanets: Can Darwin produce false positive detections?, *Astron. Astrophys.*, 388, 985-1003.

Selsis, F., A. Commeyras, M. Dobrijevic, and H. Martin (2002b) Atmospheric levels of $NO_x$ and $O_2$ on the prebiotic Earth and their possible role in the origin of life, paper presented at the 34th COSPAR scientific assembly, 10-19 October, Houston, TX, USA.

Shindell, D.T., J.L. Grenfell, D. Rind, V. Grewe, and C. Price (2001) Chemistry-climate interactions in the Goddard Institute for Space Studies general circulation model: 1. Tropospheric chemistry model description and evaluation, *J. Geophys. Res.* 106, 8047-8076.

Sinnhuber, M., J. P. Burrows, M. P. Chipperfield, C. H. Jackman, M. B. Kallenrode, K. F. Künzi, and M. Quack (2003) A model study of the impact of magnetic field structure on atmospheric composition during solar proton events, *Geophys. Res. Lett.* 30, 15, 1818.

Smith, D. S., J. Scalo, and J. C. Wheeler (2004) Transport of ionizing radiation in terrestrial-like exoplanet atmospheres, *Icarus* 171, 229-253.

Stamnes, K., Z. Jin, J. Slusser, C. Booth, T. Lucas (1992) Several-fold enhancement of biologically effective ultraviolet radiation levels at McMurdo Station, Antarctica during the 1990 ozone 'hole', *Geophys. Res. Lett.* 19, (10), 1013-1016.

Thomas, B. C., A. L. Melott, L. D. Martin, and C. H. Jackman (2004) Ozone abundance in a nitrogen-carbon-dioxide dominated terrestrial paleoatmosphere, *Astrophys., atmos. and ocean. phys.*, geophysical space series.

Tian, F., O. B. Toon, A. A. Pavlov, and H. De Sterck (2005) A hydrogen-rich early Earth atmosphere, *Science* 308, 1014-1017.

Yienger J. J. and H. Levy (1996) Empirical model of global soil-biogenic $NO_x$ emissions, *J. Geophys. Res.* 100, D6, 11447-11464.

**Tables**

Table 1: Comparison of Pre-industrial $NO_x$ sources with those on the early Earth. Unless otherwise stated, units are Teragrammes Nitrogen (Tg N) per year. E is the Extreme UV factor.

| *Mechanism* | *Reference* | *Pre-Industrial value* | *assumed Proterozoic value* |
|---|---|---|---|
| Thunderstorm Lightning | (Price and Rind, 1994) (Navarro-González *et al*., 2001) | 1.0-12.0 | 1.0-10.0 |
| Volcanic lightning | (Mvondo *et al*., 2001) | Low | 1.0-10.0 |
| Cosmic Rays | (Rozanov *et al*., 2004) (Langematz *et al*., 2005) | 3-20 ppbv increase in the middle atmosphere | (3-20)×E ppbv |
| Soil microbes | (Yienger and Levy, 1996) | 5.5 | 5.5-11.0 |

Table 2: Contribution of the photolysis rate to a particular wavelength region

(in the UV-A, UV-B and UV-C). Values represent the product ($\sigma \times \phi$ ) for a particular species, where $\sigma$=absorption cross-section, $\phi$=quantum yield, shown as the %contribution over the entire (UV-A, B, C) range. Source: DeMore et al. (1994) (to be consistent with FACSIMILE).

| ***Photolysis Reaction*** | ***%UV-A (315-400nm)*** | ***%UV-B (280-315nm)*** | ***%UV-C (200-280nm)*** |
|---|---|---|---|
| $CH_3NO_3$-->$CH_3O$+$NO_2$ | 1.4 | 51.8 | 46.8 |
| $CH_3OOH$-->$CH_3O$+OH | 0.6 | 3.3 | 96.1 |
| $H_2O_2$-->2OH | 0.3 | 1.9 | 97.8 |
| $HNO_3$-->OH+$NO_2$ | 0 | 0.1 | 99.9 |
| HONO-->OH+NO | 0.2 | 10.5 | 89.3 |
| $NO_2$-->NO+O | 67.3 | 6.9 | 25.8 |
| $O_3$-->$O_2$+O($^1$D) | 0 | 7.1 | 92.9 |
| $O_3$-->$O_2$+O($^3$P) | 0.5 | 4.7 | 94.8 |

Table 3: Overview of source gas concentrations in the thirty-six sensitivity runs.

| **Run number** | | **$H_2$** | **CO** | **$CH_4$** | **$NO_2$** |
|---|---|---|---|---|---|
| **1%$O_2$** | **10%$O_2$** | | | | |
| 1 | 19 | 1 $10^{-4}$ | 5.5 $10^{-6}$ | 1.0 $10^{-4}$ | 0.14 $10^{-9}$ |
| 2 | 20 | 1 $10^{-3}$ | 5.5 $10^{-6}$ | 1.0 $10^{-4}$ | 0.14 $10^{-9}$ |
| 3 | 21 | 5 $10^{-7}$ | 5.5 $10^{-6}$ | 1.0 $10^{-4}$ | 0.14 $10^{-9}$ |
| 4 | 22 | 1 $10^{-4}$ | 5.5 $10^{-5}$ | 1.0 $10^{-4}$ | 0.14 $10^{-9}$ |

| **Run number** | | **$H_2$** | **CO** | **$CH_4$** | **$NO_2$** |
|---|---|---|---|---|---|
| **1%$O_2$** | **10%$O_2$** | | | | |
| 5 | 23 | $1\ 10^{-4}$ | $2.7\ 10^{-8}$ | $1.0\ 10^{-4}$ | $0.14\ 10^{-9}$ |
| 6 | 24 | $1\ 10^{-4}$ | $5.5\ 10^{-6}$ | $3.0\ 10^{-4}$ | $0.14\ 10^{-9}$ |
| 7 | 25 | $1\ 10^{-4}$ | $5.5\ 10^{-6}$ | $1.0\ 10^{-4}$ | $1.53\ 10^{-9}$ |
| 8 | 26 | $1\ 10^{-3}$ | $5.5\ 10^{-6}$ | $1.0\ 10^{-4}$ | $1.53\ 10^{-9}$ |
| 9 | 27 | $5\ 10^{-7}$ | $5.5\ 10^{-6}$ | $1.0\ 10^{-4}$ | $1.53\ 10^{-9}$ |
| 10 | 28 | $1\ 10^{-4}$ | $5.5\ 10^{-5}$ | $1.0\ 10^{-4}$ | $1.53\ 10^{-9}$ |
| 11 | 29 | $1\ 10^{-4}$ | $2.7\ 10^{-8}$ | $1.0\ 10^{-4}$ | $1.53\ 10^{-9}$ |
| 12 | 30 | $1\ 10^{-4}$ | $5.5\ 10^{-6}$ | $3.0\ 10^{-4}$ | $1.53\ 10^{-9}$ |
| 13 | 31 | $1\ 10^{-4}$ | $5.5\ 10^{-6}$ | $1.0\ 10^{-4}$ | $20.0\ 10^{-9}$ |
| 14 | 32 | $1\ 10^{-3}$ | $5.5\ 10^{-6}$ | $1.0\ 10^{-4}$ | $20.0\ 10^{-9}$ |
| 15 | 33 | $5\ 10^{-7}$ | $5.5\ 10^{-6}$ | $1.0\ 10^{-4}$ | $20.0\ 10^{-9}$ |
| 16 | 34 | $1\ 10^{-4}$ | $5.5\ 10^{-5}$ | $1.0\ 10^{-4}$ | $20.0\ 10^{-9}$ |
| 17 | 35 | $1\ 10^{-4}$ | $2.7\ 10^{-8}$ | $1.0\ 10^{-4}$ | $20.0\ 10^{-9}$ |
| 18 | 36 | $1\ 10^{-4}$ | $5.5\ 10^{-6}$ | $3.0\ 10^{-4}$ | $20.0\ 10^{-9}$ |

**Figure Legends**

Figure 1: Geological Timescale in 1 $10^9$ years or Gigayears (Gyrs) summarising the development of the Earth's atmosphere. In capitals are shown the various geological periods or “eons”. In boxes are shown important events which influenced the atmosphere.

Figure 2a: Surface ozone concentration in mixing ratio (1 $10^6$) for $O_2$ (runs 1-18) set to 1% of its present atmospheric level (PAL) during the Proterozoic. Midnight values of converged runs are shown in black, midday in white.

Figure 2b: As for 2(a) but for 10% present atmospheric level (PAL) $O_2$ (runs 19-36).

Figure 3a: As for 2(a) but for OH in molecules $cm^{-3}$.

Figure 3b: As for 2(b) but for OH in molecules $cm^{-3}$.

Figure 4a: Major sources and sinks affecting OH in molecules $cm^{-3}$ $s^{-1}$ for the control run output hourly on the last day of the run. Shown is the source $O(^1D)+H_2O$-->2OH (continuous line) and the four sinks: $OH+NO_2+M$-->$HNO_3+M$ (open squares), $OH+CH_4$-->$CH_3+H_2O$ (dashed line), $OH+H_2+O_2$-->$HO_2+H_2O$ (open diamonds) and $CO+OH$-->$CO_2+H$ (filled squares).

Figure 4b: As for 4a but for the early Earth (run 19) . This run features mean Proterozoic conditions i.e. the long-lived source-gases in the middle of their estimated ranges.

# Figures

Figure 1

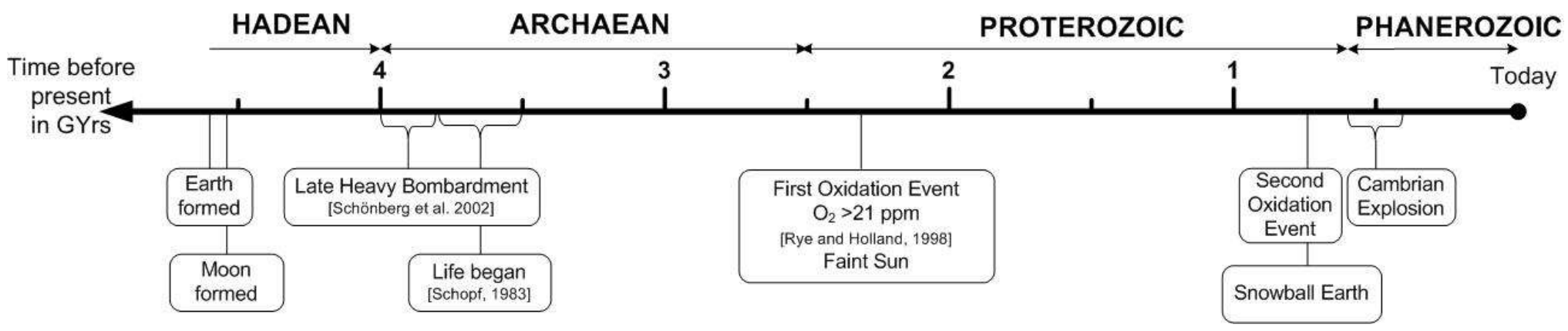

Figure 2a

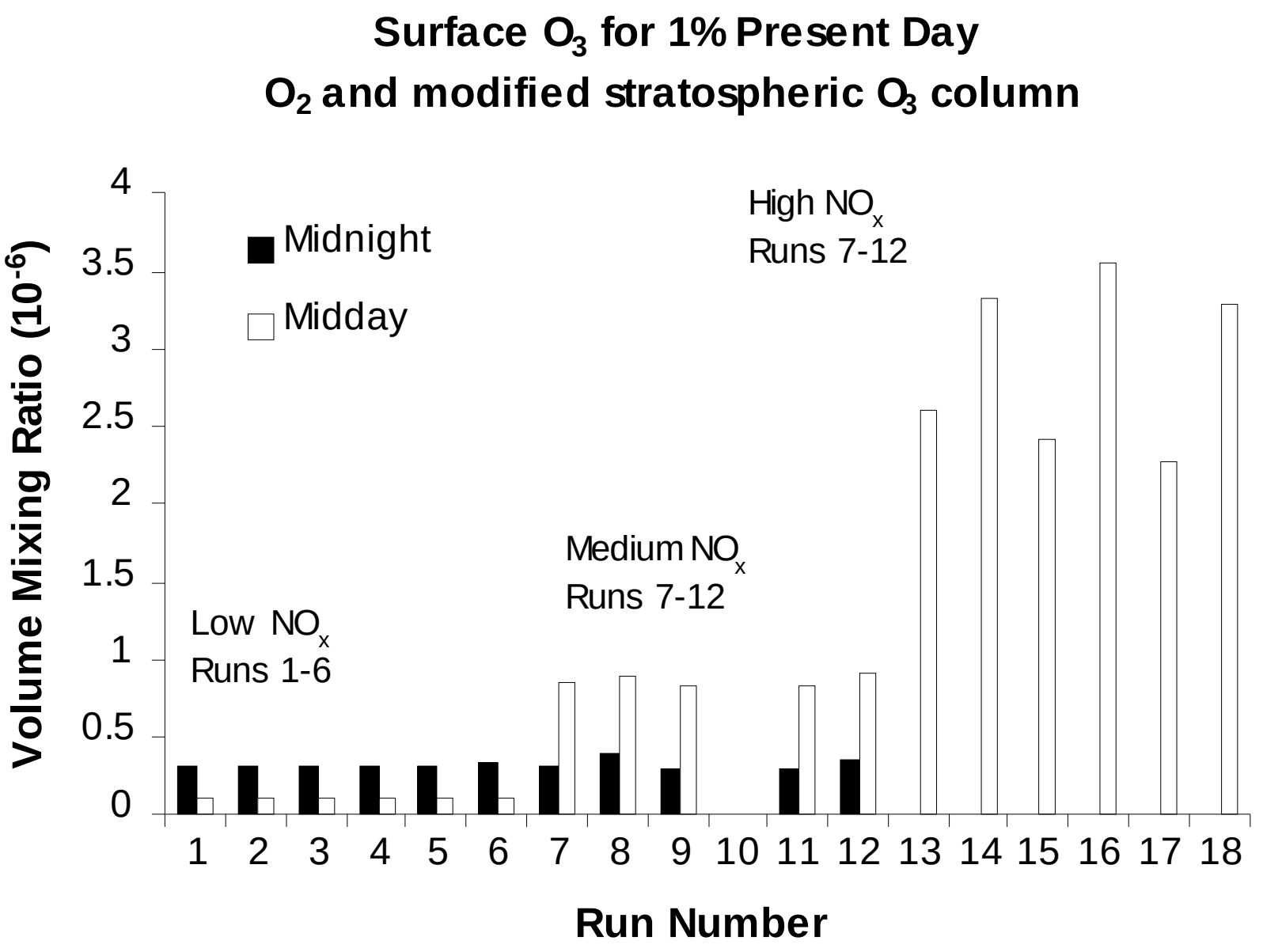

Figure 2b

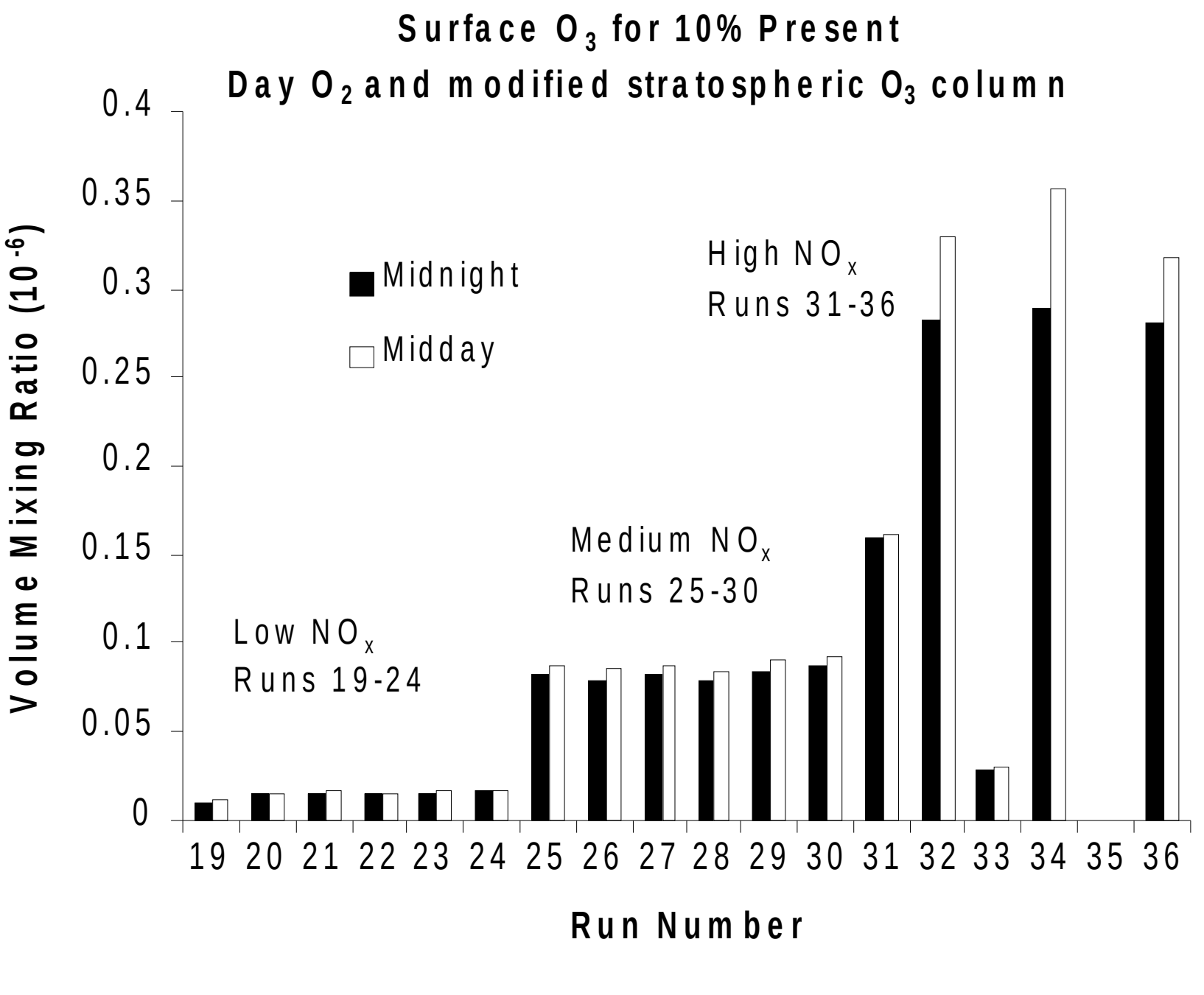

Figure 3a

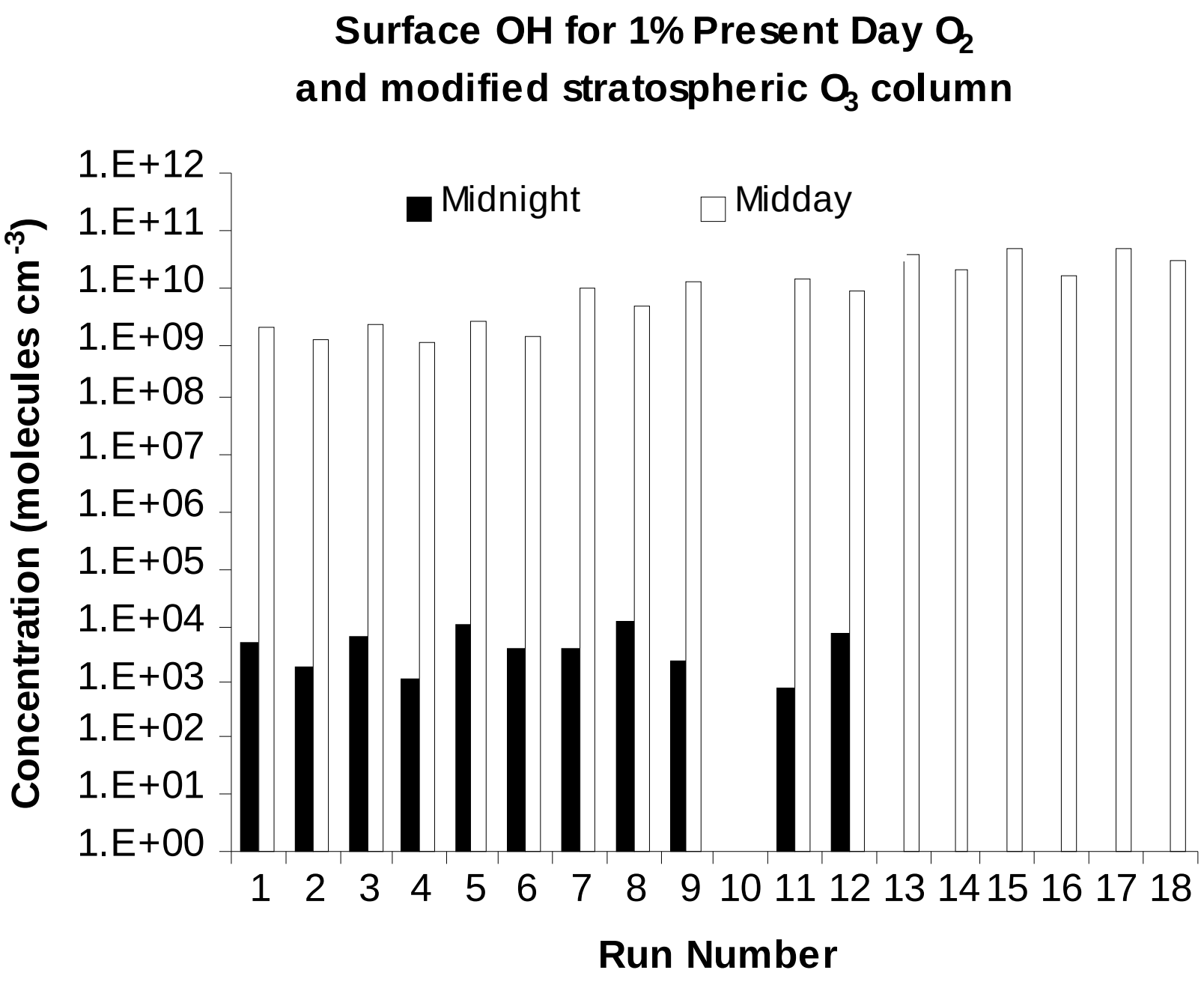

Figure 3b

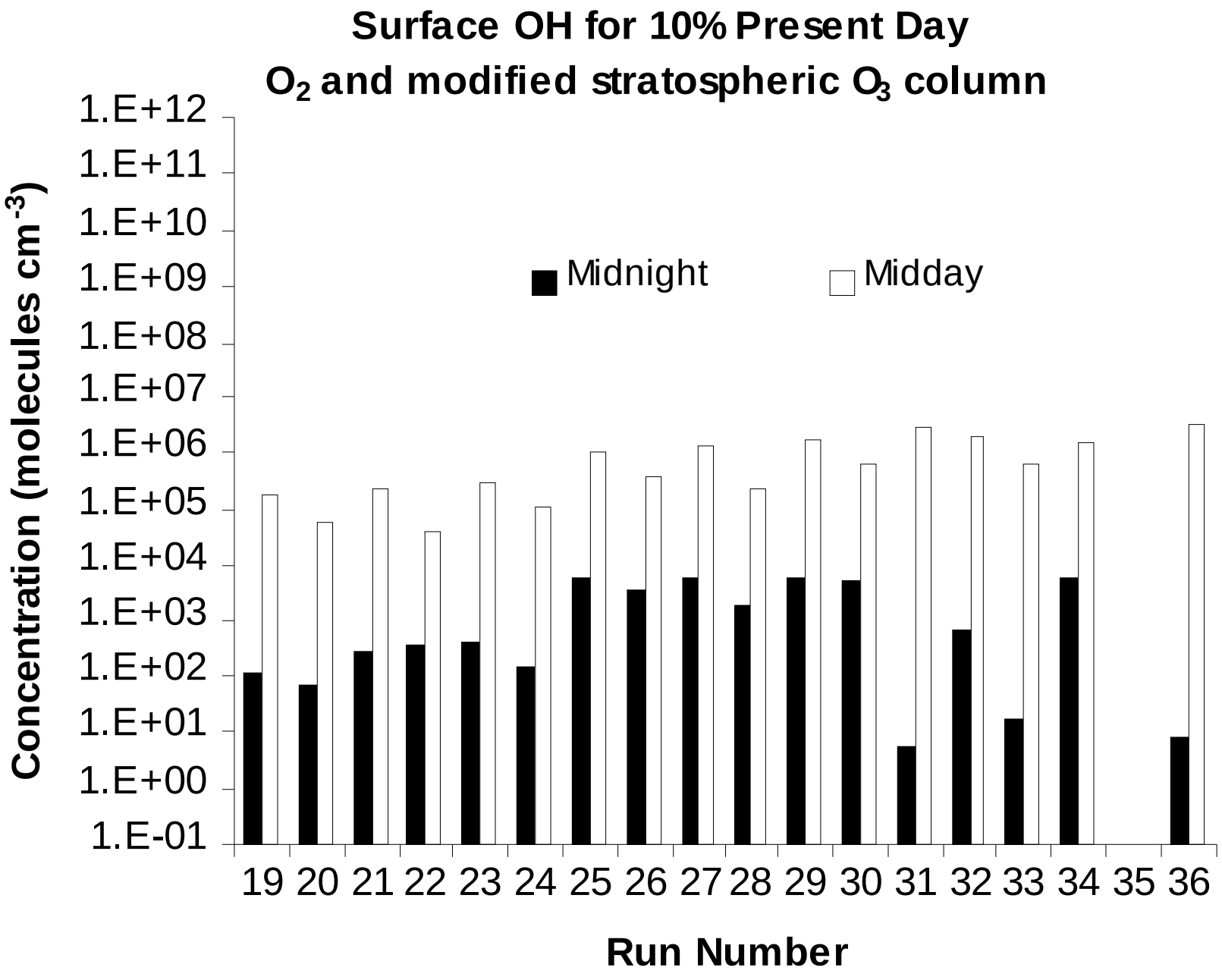

Figure 4a

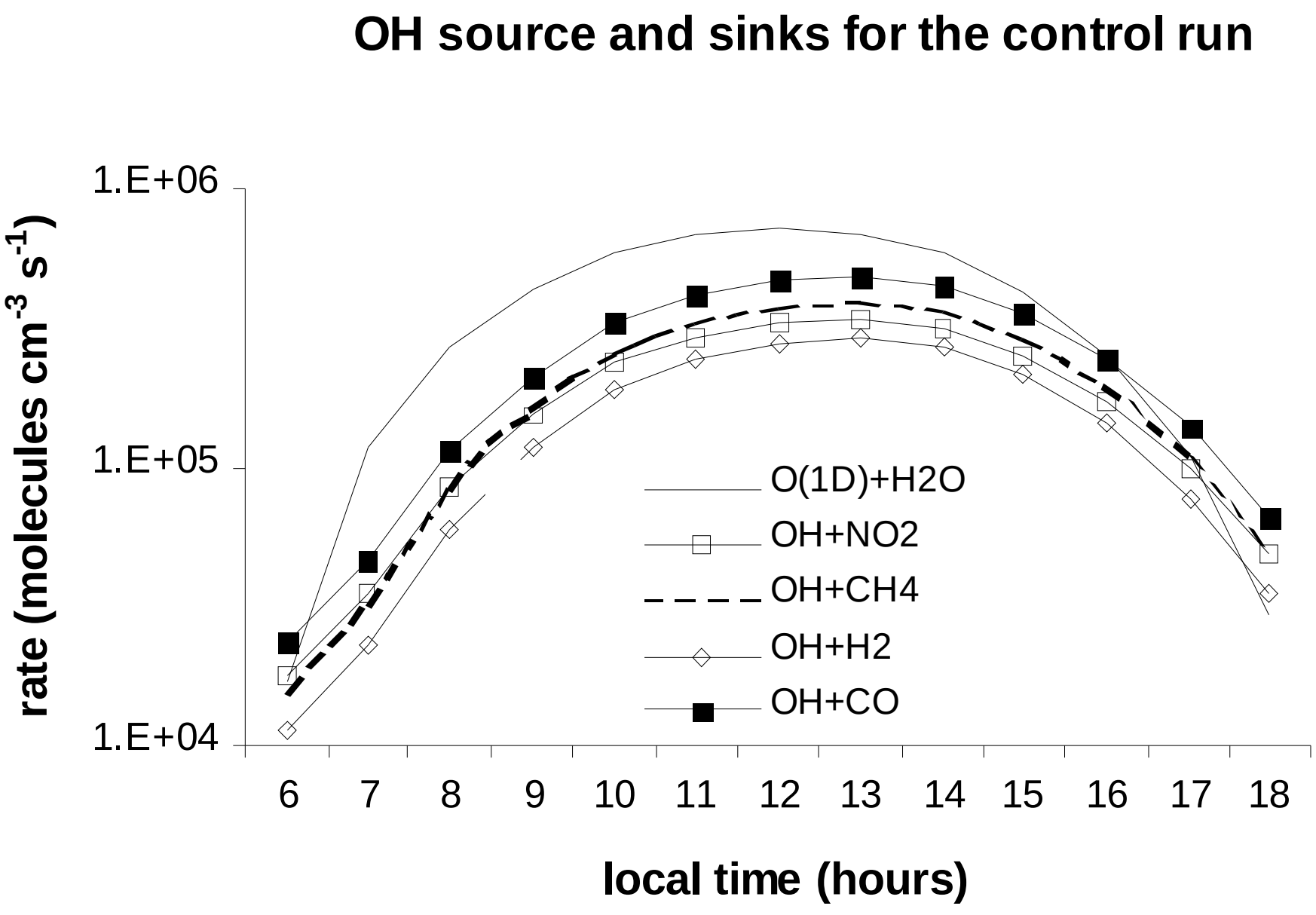

Figure 4b

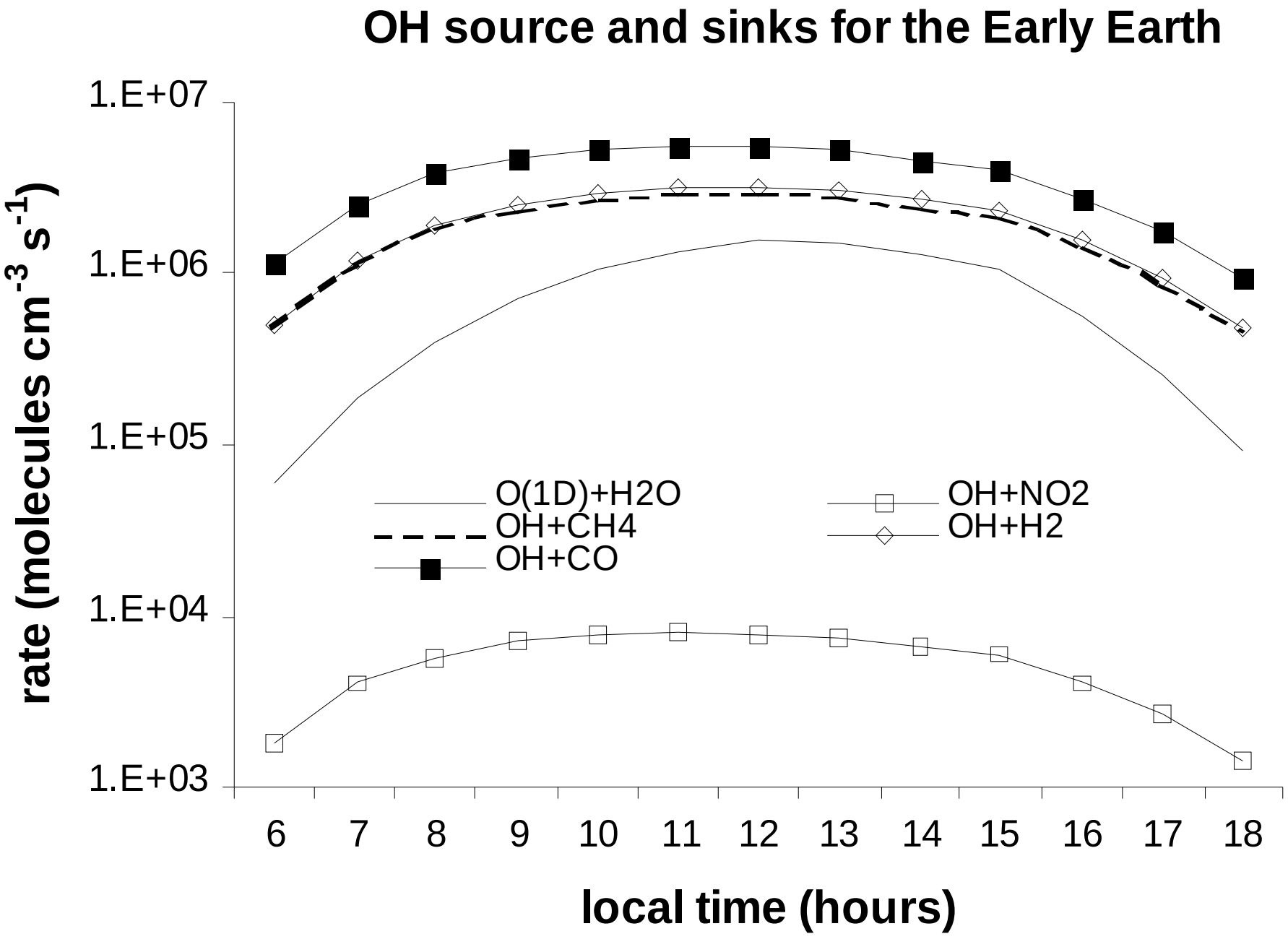